\documentclass[aps,prd,reprint,graphics,floatfix,nofootinbib,tightenlines,superscriptaddress,nobibnotes]{revtex4-2}

\usepackage{graphicx}
\usepackage[english]{babel}
\usepackage[utf8]{inputenc}
\usepackage{amsmath,amssymb}
\usepackage{amsfonts}
\usepackage{multirow}
\usepackage{pstricks}
\usepackage{float}
\usepackage{subfigure}
\usepackage{color}
\usepackage{epsfig}
\usepackage{slashed}
\usepackage[colorlinks=true, linkcolor=blue, citecolor=blue, urlcolor=blue]{hyperref}

\begin{document}

\title{Production of doubly heavy quarkonium associated with two heavy quarks via top quark decays}

\author{Juan-Juan Niu}
\email{niujj@gxnu.edu.cn}
\affiliation{Department of Physics, Guangxi Normal University,Guilin 541004, People's Republic of China}

\author{Xu-Chang Zheng}
\email{xczheng@cqu.edu.cn}
\affiliation{Department of Physics, Chongqing Key Laboratory for Strongly Coupled Physics, Chongqing University, Chongqing 401331, P.R. China}

\author{Hong-Hao Ma$^{1,}$}
\email{honghao.ma@unesp.br, corresponding author}
\affiliation{Instituto de F$\acute{\imath}$sica Te$\acute{o}$rica, Universidade Estadual Paulista,
Rua Dr. Bento Teobaldo Ferraz, 271- Bloco II, 01140-070 S$\tilde{a}$o Paulo, SP, Brazil}

\date{\today}

\begin{abstract}
In this paper, we analyze the $1 \rightarrow 4$ decay channel for the production of doubly heavy quarkonium, $(b\bar{c})$ or $(c\bar{c})$, via top-quark decays, $t \to (b\bar{c}) + c + c + \bar{s}$ and $t \to (c\bar{c}) + b + c + \bar{s}$, within the framework of nonrelativistic QCD (NRQCD).
The dominant contributions are considered in color-singlet S-wave states, i.e., $(b\bar{c})[^1S_0]$, $(b\bar{c})[^3S_1]$, $(c\bar{c})[^1S_0]$, and $(c\bar{c})[^3S_1]$. 
Our calculations show that the decay widths for $\bar{B_{c}}$, $\bar{B_{c}^{*}}$, $\eta_{c}$ and $J/\psi$  production are 0.2251, 0.3099, 0.0537 and 0.0555 MeV, respectively, resulting in ${\cal O}(10^{4}\text{--}10^{6})$ level of $\bar{B}_c^{(*)}$ events and ${\cal O}(10^{3}\text{--}10^{5})$ level of charmonium produced at LHC per year. 
In particular, we find that the dominant contribution to $\eta_{c}$ and $J/\psi$ production via top-quark decays arises from this decay channel proposed in this work. 
Moreover, this multi-body top-quark decay process can serve as a sensitive probe for validating the narrow-width approximation (NWA).
Finally, we provide a detailed analysis of theoretical uncertainties and differential distributions to facilitate the corresponding experimental searches.
The production of a hadron associated with three quarks contains rich physical information, providing new insights for the LHC to study $B_c$ mesons and charmonia.

\end{abstract}

\maketitle

\noindent\textbf{\textit{Introduction}---}Doubly heavy mesons play a special role in heavy-flavor physics because their production and decay involve QCD interactions in both the perturbative and non-perturbative regimes. Among them, charmonium ($\eta_c$ and $J/\psi$) and bottomonium ($\eta_b$ and $\Upsilon$) are hidden-flavor bound states, whereas the $B_c$ meson is the only meson composed of two distinct heavy quarks. Historically, charmonium opened the modern era of heavy-flavor physics with the discovery of $J/\psi$ in 1974 \cite{E598:1974sol, SLAC-SP-017:1974ind}, while the $B_c$ meson was first observed much later by the CDF Collaboration at the Tevatron in 1998 \cite{CDF:1998axz}. These heavy systems provide an important laboratory for heavy-flavor phenomenology, and their unique properties also offer a distinctive window for studying CP violation. Nonrelativistic QCD (NRQCD) \cite{Bodwin:1994jh, Petrelli:1997ge} and fragmentation functions \cite{Braaten:1993jn, Zheng:2019gnb} provide effective theoretical frameworks for studying the production, decay, and interactions of such heavy mesons and quarkonia.
 
The production mechanisms of these mesons has been comprehensively studied, including
the direct hadronic production \cite{Chang:1992jb, Chang:1994aw, Gershtein:1994jw, Berezhnoy:1994ba, Chang:2003cr, Chang:2004bh, Chang:2005bf}, $e^+e^-$ production \cite{Zheng:2015ixa, Zhang:2021ypo, Yang:2022zpc, Zhan:2022etq}, photoproduction \cite{Chen:2014frw, Sun:2015hhv, Bi:2016vbt, Zhan:2022nck, Cai:2026hll}, and heavy-ion collisions \cite{Chen:2018obq}, but also their indirect production channels via top-quark decay \cite{Qiao:1996rd, Chang:2007si, Wu:2008cn, Sun:2010rw}, $Z^0$-boson \cite{Chang:1992bb, Qiao:2011zc, Deng:2010aq, Yang:2010yg, Wang:2023ssg, Zheng:2022mds}, $W^+$-boson \cite{Qiao:2011yk, Liao:2011kd, Liao:2012rh, Zheng:2019egj} and Higgs-boson \cite{Jiang:2015pah, Zheng:2023atb} decays. At high-energy colliders, top factories provide a unique perspective for the study of mesons. Early studies of $B_c$ and its excited states through top quark decays were conducted in process $t \to (b\bar{c}) + c + W^+$ \cite{Qiao:1996rd, Chang:2007si}, and by including the rare flavor changing neutral currents (FCNC) process $t \to (c\bar{Q}) + Q + Z^0$ \cite{Niu:2018tvo} to investigate the semi-inclusive production of charmonium and $B_c$ mesons. An Event generator for hadronic production of the $B_c$ meson has been developed, called BCVEGPY \cite{Chang:2003cq, Chang:2005hq}

Utilizing LHC Run 3 data at $\sqrt{s} = 13.6\text{ TeV}$, the ATLAS Collaboration has measured the $t\bar{t}$ production cross section to be $850 \pm 27\text{ pb}$, which is consistent with the Standard Model prediction of $924^{+32}_{-40}\text{ pb}$~\cite{ParticleDataGroup:2024cfk}.
The abundant yields of top quarks motivates us to study their multi-body decay processes.
The production of a hadron associated with three quarks contains rich physical information, providing a new approach for the LHC to detect $B_c$ mesons and charmoniums. In this paper, we study the production of $B_c$ mesons and charmoniums through four-body top-quark decay channels, $t \to (b\bar{c}) + c + c + \bar{s}$ and $t \to (c\bar{c}) + b + c + \bar{s}$, within the NRQCD factorization framework. Moreover, the event yields at facilities, such as LHC, CEPC, and LHeC can be estimated for the production of $B_c^{(*)}$ and charmoniums associated with three (anti-)quarks, which provides an indirect but potentially efficient source of heavy mesons with kinematics characteristics determined by the mass of heavy quarks and multi-body phase space. However, this $1\to4$ decay process is relatively complicated, and its phase space integration requires elaborate reduction procedures.

It is also instructive to examine the narrow-width approximation (NWA)~\cite{Breit:1936zzb,Jackson:1975vf} within this manuscript. NWA is widely utilized for intermediate $W$-boson process and is generally robust given $\Gamma_W/M_W \ll 1$, allowing for a straightforward factorization of the associated production and subsequent decay processes. However, as dedicated studies have highlighted, off-shell and non-factorizable effects can become prominent in restricted phase-space regions, particularly near kinematic endpoints~\cite{Uhlemann:2009,Heinrich:2018}. Because the $1 \to 4$ top-quark decay channels explored here feature complex multi-body kinematics and nontrivial correlations, our full phase-space calculation provides not only exact theoretical predictions but also a rigorous quantitative benchmark to assess the validity of the NWA for $W$-mediated heavy-quarkonium production.

\begin{figure}[htp]
\centering
\subfigure[]{\includegraphics[width=0.11\textwidth]{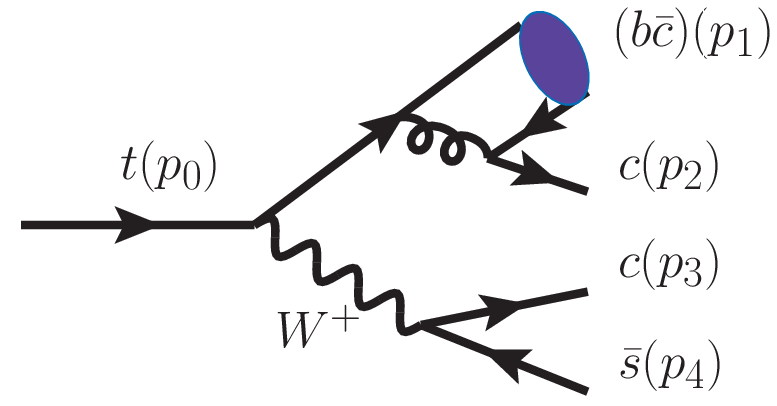}}
\subfigure[]{\includegraphics[width=0.11\textwidth]{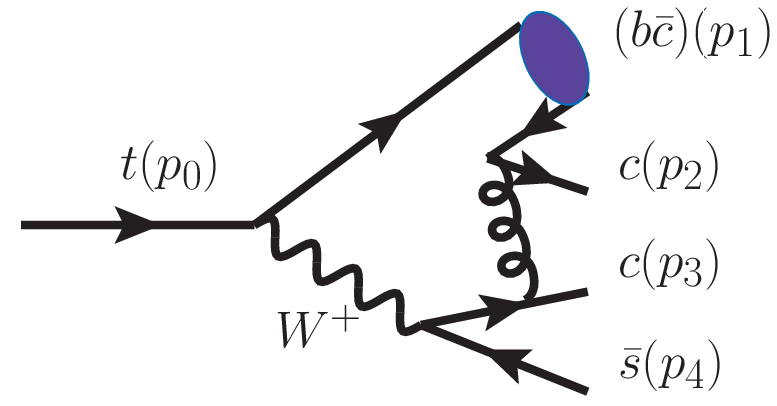}}
\subfigure[]{\includegraphics[width=0.11\textwidth]{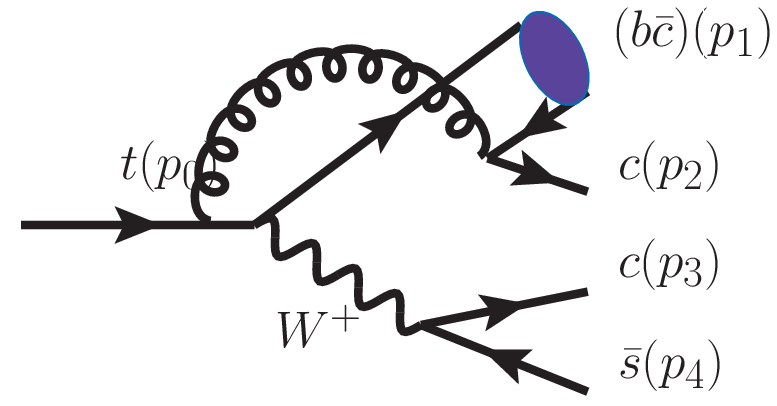}}
\subfigure[]{\includegraphics[width=0.11\textwidth]{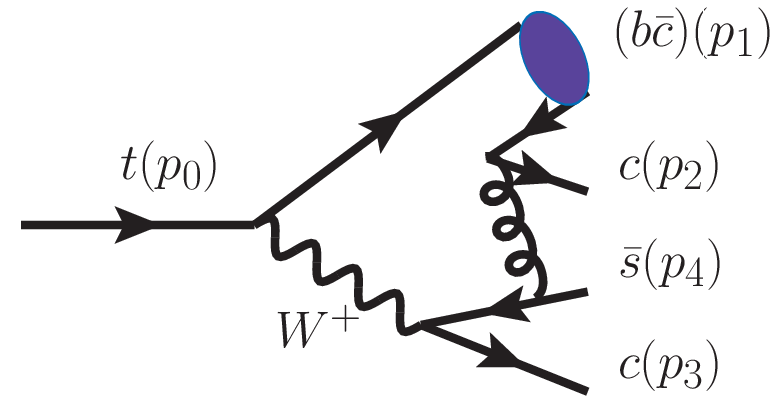}}\\
\subfigure[]{\includegraphics[width=0.11\textwidth]{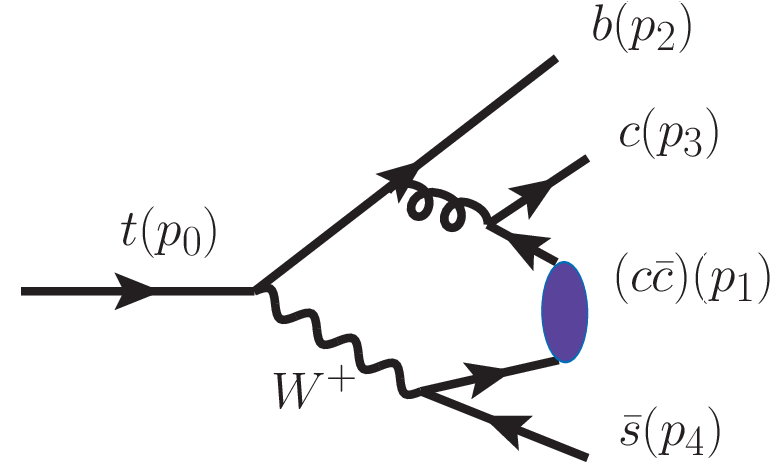}}
\subfigure[]{\includegraphics[width=0.11\textwidth]{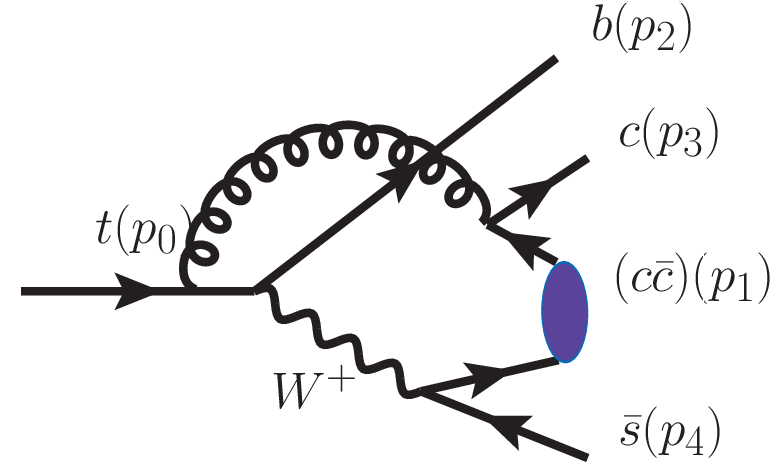}}
\subfigure[]{\includegraphics[width=0.11\textwidth]{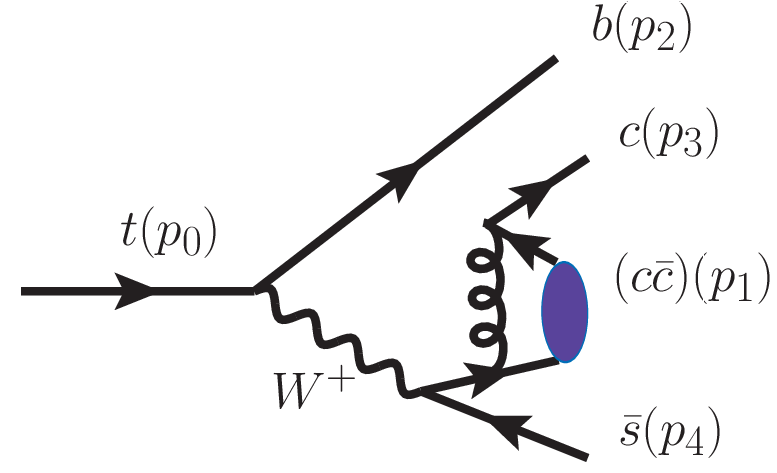}}
\subfigure[]{\includegraphics[width=0.11\textwidth]{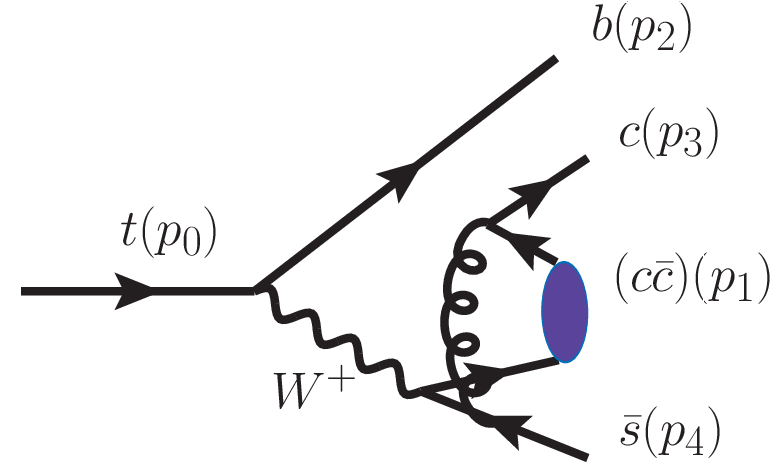}}
\caption{Typical Feynman diagrams for $t(p_0)\to (b\bar{c})(p_1)+c(p_2)+ c(p_3)+ \bar{s}$ and $t(p_0)\to (c\bar{c})(p_1)+b(p_2)+ c(p_3)+ \bar{s}$, where $b\bar{c}$ and $c\bar{c}$ quarkonium are in color-singlet states, i.e. $(^1S_0)_1$ and $(^3S_1)_1$, respectively.} \label{feyn}
\end{figure}

\noindent\textbf{\textit{Calculation~Techniques}.---}Typical Feynman diagrams for the process $t(p_0)\to (b\bar{c})(p_1)+c(p_2)+c(p_3)+\bar{s}(p_4)$ are shown in Fig.~\ref{feyn}(a--d), and the other four diagrams can be obtained by exchanging the two identical charm quarks in the final state; Four Feynman diagrams for $t(p_0)\to (c\bar{c})(p_1)+b(p_2)+c(p_3)+\bar{s}(p_4)$ are shown in Fig.~\ref{feyn}(e--h)

Within the NRQCD framework \cite{Bodwin:1994jh, Petrelli:1997ge}, the  decay width for the production of $B_c$ meson (as a representative) can be factorized as
\begin{eqnarray}
\Gamma(t && \rightarrow \bar{B}_{c}^{(*)} + c + c + \bar{s}) \nonumber\\
&&= \sum_{n} \hat{\Gamma}(t \rightarrow  \langle b\bar{c}\rangle [n] + c + c + \bar{s}) \times \langle \mathcal{O}^H [n] \rangle,
\end{eqnarray}
where $[n]$ labels the spin and color quantum number of the intermediate $\langle b\bar{c}\rangle$-diquark state, such as $[^1S_0]_{1}$, $[^3S_1]_{1}$. $\langle \mathcal{O}^H[n] \rangle$ denotes the non-perturbative long-distance matrix element from the $\langle b\bar{c} [n] \rangle$-diquark to the quarkonium $\bar{B}_{c}^{(*)}$, which is proportional to the transition probability and can be related to the Schrödinger wave function at the origin $|\Psi_{b\bar{c}}(0)|$ for $S$-wave states. The decay width $\hat{\Gamma}((t \rightarrow  (b\bar{c}) [n] + c + c + \bar{s})$ is the perturbative short-distance coefficients and can be expressed as

$$ \hat{\Gamma}(t \rightarrow  (b\bar{c}) [n] + c + c + \bar{s}) = \int \frac{1}{2m_t} \overline{\sum} |\mathcal{M}[n]|^2 d\Phi_4,$$
where $\mathcal{M}$ stands for the hard amplitude, $\overline{\sum}$ is to sum over the colors and spins of all final-state particles ($(b\bar{c})$, $c$, $c$ and $\bar{s}$) and to average over the spin and color of the initial top quark. $d\Phi_4$ is the four-body phase space and can be split into two lower-dimensional sub-systems by inserting the following identities:
\begin{align*}
\int \frac{d^4 p_t}{(2\pi)^4} (2\pi)^4 \delta^4(p_t - p_2 - p_3) &= 1, \\
s \int \frac{dy}{2\pi} 2\pi \delta(p_t^2 - sy) &= 1,
\end{align*}
where $sy$ denotes the invariant mass squared of the $(p_2, p_3)$ system ($sy=p_t^2 = s_{23}=(p_2+p_3)^2$).
Then $d\Phi_4$ can be rewritten as,
\begin{align*}
d\Phi_4 &= (2\pi)^4 \delta^4 \left(p_0 -\sum_{i=1}^4 p_i\right) \prod_{i=1}^{4} \frac{d^3 p_i}{(2\pi)^3 2p_i^0} \nonumber \\
&= s \int \frac{dy}{2\pi} \theta(y) d\Phi_3(p_0; p_1, p_t, p_4) d\Phi_2(p_t; p_2, p_3).
\end{align*}
Firstly, we evaluate $d\Phi_3$ in the rest frame of $p_0$, where $p_0 = (\sqrt{s}, 0, 0, 0)$. To simplify the angular integration, we define the following coordinate system:
\begin{itemize}
    \item $p_1 = (p_1^0, 0, 0, |\vec{p}_1|)$,~aligned with the $z$-axis.
    \item $p_4 = (p_4^0, |\vec{p}_4|\sin\theta_4, 0, |\vec{p}_4|\cos\theta_4)$,~in the $x$-$z$ plane.
    \item $\vec{p}_t = -(\vec{p}_1 + \vec{p}_4)$,~momentum conservation.
\end{itemize}
Then $d\Phi_3$ can be simplified as:
\begin{equation}
d\Phi_3 = \frac{1}{4(2\pi)^3} dp_1^0 dp_4^0.
\end{equation}
However the 2-body phase space $d\Phi_2$ is most easily evaluated in the rest frame of $p_t$:
\begin{equation}
d\Phi_2(p_t; p_2, p_3) = \frac{|\vec{p}_2^{~\prime}|}{4(2\pi)^2 \sqrt{sy}} d\cos\theta_2' d\phi_2',
\end{equation}
where $|\vec{p}_2^{~\prime}|$ is the magnitude of the 3-momentum of particle 2 in the $p_t$ rest frame, $\theta_2^{\prime}$ and $\phi_2^{\prime}$ are the azimuth angle at this rest frame.
Combining all components, the complete differential 4-body phase space is:
\begin{equation}
d\Phi_4= \frac{s}{2^4(2\pi)^6} \frac{|\vec{p}_2^{~\prime}|}{\sqrt{sy}} dy \, d p_1^0 d p_4^0 \, d\cos\theta_2^{\prime} d\phi_2^{\prime}.
\end{equation}
The amplitude $\mathcal{M}$ is obtained by summing up all tree-level diagrams for $t\to b+W^+$, $W^+\to c\bar{s}$ with an additional gluon splitting that produces a hard $c\bar{c}$ quark pair.   Then the quarkonium is produced by projecting an on-shell $b\bar{c}$ or $c\bar{c}$ pair onto the desired spin and color configurations $(b\bar{c})[n]$ or $(c\bar{c})[n]$.
For example, the amplitude of Feynman diagram Fig. \ref{feyn}(a) can be written as
\begin{equation}
\mathcal{M}_a = i \mathcal{K}_a \cdot L_{\mu} \cdot H^{\mu},
\end{equation}
with
\begin{eqnarray}
L_{\mu}&=&\bar{u}_c(p_3) \gamma_\mu (1-\gamma^5) v_s(p_4),\nonumber\\
H^{\mu}&=& \bar{u}_c(p_2) \gamma^\rho \Pi(p_1) \gamma_\rho \frac{\slashed{p}_1 + \slashed{p}_2 + m_b}{(p_1 + p_2)^2 - m_b^2} \gamma^\mu (1-\gamma^5) u_t(p_0),\nonumber
\end{eqnarray}
where $\mathcal{K}_a$ is the overall factor including the color, coupling constant and propagators.
$p_1=p_{11}+p_{12}$, where $p_{11}$ and $p_{12}$ are the momenta of these two constituent quarks ($b$ and $\bar{c}$ quarks) of $(b\bar{c})$-quarkonium, respectively. More explicitly, we have
\begin{equation}
p_{11}=\frac{m_b}{M}p_1+q \;\;{\rm and}\;\;
p_{12}=\frac{m_c}{M}p_1-q,
\end{equation}
where $M\simeq m_b+m_c$, and $q$ is the relative momentum between two constituent quarks.  
Relating to the spin angular momentum of
$(b\bar{c})$-quarkonium, the form of the projectors can
be conveniently written as
\begin{equation}
\Pi^0_{p_1}=\frac{1} {2\sqrt{M}}\gamma_5
(\slashed{p}_1+M)\;\;,\;\; \Pi^\alpha_{p_1}=\frac{1}
{2\sqrt{M}} \gamma^\alpha(\slashed{p}_1+M).
\end{equation} 

\noindent\textbf{\textit{Numerical~Results}---}The input parameters in numerical calculation are taken as follows \cite{ParticleDataGroup:2024cfk,Eichten:1994gt,Eichten:1995ch}
\begin{eqnarray} \label{para}
&&|R_{b\bar{c}}(0)| = 1.642 \, \mathrm{GeV^{3/2}},  ~~~~|R_{c\bar{c}}(0)| = 0.801 \, \mathrm{GeV^{3/2}},  \nonumber\\
&&\sin^2 \theta_W = 0.231,  ~\Gamma_{W}=2.14~\mathrm{GeV},~~m_t=172.5\, \mathrm{GeV},\nonumber\\
&&m_b = 4.8 \, \mathrm{GeV}, ~~~~m_c = 1.5 \, \mathrm{GeV}, ~~~~~M_{bc}=m_b+m_c, \nonumber\\
&&M_{cc}=m_c+m_c, ~~\alpha = 1/128, ~~~~~~~\alpha_s=0.237.
\end{eqnarray}
The decay widths for the production of $\bar{B}_c$, $\bar{B}_c^*$ mesons and $\eta_c$, $J/\psi$ quarkonium through top quark decays are obtained, and the results are
\begin{eqnarray}
&&\Gamma_{t \to \bar{B}_c + c + c + \bar{s}}=0.2251~\mathrm{MeV},\\
&&\Gamma_{t \to \bar{B}_c^{*} + c + c + \bar{s}}=0.3099~\mathrm{MeV},\\
&&\Gamma_{t \to \eta_c + b + c + \bar{s}}=0.0537~\mathrm{MeV}, \\
&&\Gamma_{t \to J/\psi + b + c + \bar{s}}=0.0555~\mathrm{MeV}.
\end{eqnarray}

The results show that the decay width for the production of $\bar{B}_c^{*}$ is approximately 1.33 times larger than that of $\bar{B}_c$, while the decay width of $\eta_c$ is slightly smaller than that of $J/\psi$, and they are of comparable magnitude. The $(b\bar{c})$ yield comes almost entirely from Feynman diagrams (a) and (c) in Fig.~\ref{feyn}, indicating that $B_c$ mesons are mainly produced through the fragmentation of the $b$ jet and the top quark. The contribution from (b) and (d) in Fig.~\ref{feyn}, where the split gluon originates from the products of $W^+$ decay, is about three orders of magnitude smaller. 
Consequently, in this channel we cannot distinguish whether the produced $(b\bar{c})$ is prompt or non-prompt.
In contrast, the $(c\bar{c})$ yield is almost entirely from diagrams (g) and (h), i.e., from a sub-process like $W^+\to (c\bar{c})\, c\,\bar{s}$. 
Then, the produced $(c\bar{c})$ should be considered as prompt.
The main reason is that in the top quark decays, the $b$ quark and $W^{+}$ boson are typically back-to-back with a large opening angle, which makes it difficult for quarks from their decays to ``meet'' in phase space.  
In other words, the process $t \to (b\bar{c}) + c + c + \bar{s}$ can be viewed as $t \to (b\bar{c}) + c + W^{+}$ $\bigotimes$ $W^+ \to c + \bar{s}$, while $t \to (c\bar{c}) + b + c + \bar{s}$ can be viewed as $t \to b +W^{+}$ $\bigotimes$ $W^+ \to (c\bar{c}) + c + \bar{s}$. 

NWA relies on the assumption that the intermediate $W$ boson is on-shell. Mathematically, the squared Breit-Wigner propagator can be approximated by a Dirac delta function when $\Gamma_W / M_W \to 0$~\cite{Breit:1936zzb}:
\begin{eqnarray}
\frac{1}{(q^2_W - M_W^2)^2 + M_W^2 \Gamma_W^2} \approx \frac{\pi}{M_W \Gamma_W} \delta(q^2_W - M_W^2),   
\end{eqnarray}
where $q_W$ is the momentum of the intermediate $W$ boson. By integrating over the phase space using this equation, the total $1 \to 4$ decay width can be factorized into the production and decay of the on-shell $W$ boson. 
NWA should provide a very good approximation, i.e.,
\begin{eqnarray}
&&\Gamma_{t \to (b\bar{c}) + c + c + \bar{s}}^{\text{NWA}} \approx \Gamma_{t \to (b\bar{c}) + c + W^{+}}\times \frac{\Gamma_{W^+ \to c\bar{s}}}{\Gamma_W}, \\
&&\Gamma_{t \to (c\bar{c}) + b + c + \bar{s}}^{\text{NWA}} \approx \Gamma_{t \to b + W^{+}}\times \frac{\Gamma_{W^+ \to (c\bar{c}) + c + \bar{s}}}{\Gamma_W}.
\end{eqnarray}
Using the parameter set given in Eq.~(\ref{para}), we have calculated all the decay widths involved above.
\begin{eqnarray}
&&\Gamma_{t \to \bar{B}_c + c + W^{+}}=0.7215~\mathrm{MeV},\\
&&\Gamma_{t \to \bar{B}_c^{*} + c + W^{+}}=0.9941~\mathrm{MeV},\\
&&\Gamma_{W^{+} \to \eta_c + c + \bar{s}}=0.0794~\mathrm{MeV}, \\
&&\Gamma_{W^{+} \to J/\psi + c + \bar{s}}=0.0820~\mathrm{MeV},\\
&&\Gamma_{t \to b + W^{+}}=1.54~\mathrm{GeV},\\
&&\Gamma_{W^{+} \to c + \bar{s}}=0.7027~\mathrm{GeV},
\end{eqnarray}
which implies
\begin{eqnarray}
&&\Gamma_{t \to \bar{B}_c + c + c + \bar{s}}^{\text{NWA}} \approx 0.2369~\mathrm{MeV},\\
&&\Gamma_{t \to \bar{B}_c^{*} + c + c + \bar{s}}^{\text{NWA}}\approx 0.3264~\mathrm{MeV},\\
&&\Gamma_{t \to \eta_c + b + c + \bar{s}}^{\text{NWA}} \approx 0.0571~\mathrm{MeV}, \\
&&\Gamma_{t \to J/\psi + b + c + \bar{s}}^{\text{NWA}} \approx 0.0590~\mathrm{MeV}.
\end{eqnarray}
In our numerical calculation, a naive comparison might suggest that the NWA result is slightly larger than the full phase-space result. However, this slight enhancement is an artifact of utilizing the calculated tree-level partial width for the $W$ boson. The total decay width $\Gamma_W$ and the corresponding partial widths encompass higher-order corrections and experimental uncertainties that are not fully captured in a pure leading-order calculation. If we instead directly adopt the experimental branching fraction from the Particle Data Group (PDG), $\mathcal{B}(W^+ \to c\bar{s}) \approx 31\%$~\cite{ParticleDataGroup:2024cfk}, the NWA prediction becomes slightly smaller than the exact full phase-space result by approximately 0.6\%. This is in excellent agreement with the theoretical expectation, where the finite-width correction is estimated to be at the level of $\mathcal{O}(\Gamma_{W}/M_{W}) \approx 2.6\%$.

Then the branching ratios (Br) of these mesons are analyzed and listed in Table~\ref{tab:br_events}. The definition of the branching ratios in the table is ${\rm Br}=\Gamma/\Gamma_{t\to b+W^+}$, where the decay width of $t \to b + W^+$ can be regarded as the total decay width of the top quark and be 1.54 GeV.
To estimate the produced events ($N$) of $\bar{B}_c$, $\bar{B}_c^*$ mesons and $J/\psi$, $\eta_c$ quarkonium through this decay channel, 
we take the numbers of produced top-quark events at the LHC, CEPC and LHeC to be $10^{8\text{--}10}$ \cite{Kidonakis:2004hr, Kuhn:2013zoa}, $0.6\times10^{6}$ \cite{CEPCStudyGroup:2023quu,Ai:2024nmn} and ${\cal O}(10^{5})$ \cite{LHeC:2020van} per year, respectively. Table~\ref{tab:br_events} reveals that

\begin{itemize}
\item There are sizable expected yields at LHC due to the sufficient annual $t\bar{t}$ statistics ($10^{8\text{--}10}$), resulting in about $1.5 \times 10^{4-6}$, $2.0 \times 10^{4-6}$, $3.5 \times 10^{3-5}$ and $3.6 \times 10^{3-5}$ of $\bar{B}_c$, $\bar{B}_c^{*}$, $\eta_c$ and $J/\psi$ events produced per year correspondingly.

\item The $2S$ states of $b\bar{c}$ and $c\bar{c}$ are also non-negligible, and we adopt $|R_{b\bar{c}(2S)}(0)| = 0.983 \, \mathrm{GeV^{3/2}}$ and $|R_{c\bar{c}(2S)}(0)| = 0.529 \, \mathrm{GeV^{3/2}}$~\cite{Eichten:1995ch}. Then one can expect that there will be $5.2 \times 10^{3-5}$, $7.2 \times 10^{3-5}$, $1.5 \times 10^{3-5}$ and $1.6 \times 10^{3-5}$ of $\bar{B}_c(2S)$, $\bar{B}_c^{*}(2S)$, $\eta_c(2S)$ and $J/\psi(2S)$ events in LHC per year respectively.

\item For future $e^+e^-$ and $ep$ top factories such as CEPC and LHeC, the annual top-quark yields are at most at the ${\cal O}(10^{5})$ level. Consequently, the numbers of $b\bar{c}$ and $c\bar{c}$ quarkonium produced via top-quark decays are very limited and can be neglected.

\end{itemize}

\begin{table}[htbp]
\centering
\caption{Branching ratios and produced events of $\bar{B}_c$, $\bar{B}_c^{*}$, $J/\psi$ and $\eta_c$ at LHC, CEPC and LHeC, respectively. }
\begin{tabular}{ccccc}
\hline\hline
State & Br & $N_{\rm LHC}$ &  $N_{\rm CEPC}$ & $N_{\rm LHeC}$ \\
\hline
$\bar{B}_c$      & $1.4617\times10^{-4}$ & $1.4617\times10^{4-6}$ & $8.7701\times10^{1}$ & $1.4617\times10^{1}$\\
$\bar{B}_c^{*}$  & $2.0123\times10^{-4}$ & $2.0123\times10^{4-6}$ & $1.2074\times10^{2}$ & $2.0123\times10^{1}$\\
$\eta_c$         & $3.4870\times10^{-5}$ & $3.4870\times10^{3-5}$ & $2.0922\times10^{1}$ & $3.4870$\\
$J/\psi$         & $3.6039\times10^{-5}$ & $3.6039\times10^{3-5}$ & $2.1623\times10^{1}$ & $3.6039$\\
\hline\hline
\end{tabular}
\label{tab:br_events}
\end{table}

Overall, among these four mesons produced by this considered decay channel, $\bar{B}_c^{*}$ seems to be the most abundant state at high-energy colliders, suggesting that $\bar{B}_c^{*}\to \bar{B}_c+\gamma$ feed-down may play an important role in realistic $\bar{B}_c$ yields in top-quark decay.

Next, the theoretical uncertainties caused by the heavy-quark masses ($m_b$ and $m_c$) are analyzed for the decay widths of $\bar{B}_c$, $\bar{B}_c^{*}$, $\eta_c$ and $J/\psi$. During this discussion, we keep the central value of one heavy-quark mass fixed and analyze the change in the decay width caused by the other one, with the ranges $m_b=4.8\pm0.3\,\mathrm{GeV}$ and $m_c=1.5\pm0.1\,\mathrm{GeV}$. The relevant results are shown in Table~\ref{tab:unc_masses}, which reveals that

\begin{itemize} \label{uncer_item}
\item As $m_b$ scans over $4.5\to5.1\,\mathrm{GeV}$, the decay widths show only small variations: $\Gamma(t\to \bar{B}_c+c+c+\bar{s})=0.2251_{-0.53\%}^{+0.53\%}\,\mathrm{MeV}$ and $\Gamma(t\to \bar{B}_c^{*}+c+c+\bar{s})=0.3099_{-1.90\%}^{+1.10\%}\,\mathrm{MeV}$. This indicates a weak dependence on $m_b$, while the decay widths for $\eta_c$ and $J/\psi$ are almost independent of $m_b$. 

\item However, the results show a much stronger sensitivity to $m_c$. For $m_c=1.5\pm0.1\,\mathrm{GeV}$, the widths become $\Gamma(t\to \bar{B}_c+c+c+\bar{s})=0.2251_{-17.95\%}^{+23.81\%}\,\mathrm{MeV}$, $\Gamma(t\to \bar{B}_c^{*}+c+c+\bar{s})=0.3099_{-19.69\%}^{+25.72\%}\,\mathrm{MeV}$, $\Gamma(t\to \eta_c +b+c+\bar{s})=0.0537_{-17.88\%}^{+23.46\%}\,\mathrm{MeV}$, and $\Gamma(t\to J/\psi+b+c+\bar{s})=0.0555_{-17.84\%}^{+23.60\%}\,\mathrm{MeV}$, reflecting the stronger dependence of the multi-body phase space and short-distance amplitudes on $m_c$.
\end{itemize}

\begin{table}[htbp]
\centering
\caption{Uncertainty from the heavy quark masses ($m_b$ and $m_c$) for the decay widths (unit in MeV) of $\bar{B}_c$, $\bar{B}_c^{*}$, $\eta_c$ and $J/\psi$.}
\begin{tabular}{c|ccc|ccc}
\hline\hline
\multirow{2}{*}{State} & \multicolumn{3}{c|}{$m_b$ (GeV) } & \multicolumn{3}{c}{$m_c$ (GeV)}\\
\cline{2-7}
 & $4.5$ & $4.8$ & $5.1$ & $1.4$ & $1.5$ & $1.6$\\
\hline
$\bar{B}_c$     & $0.2263$ & $0.2251$ & $0.2239$ & $0.2787$ & $0.2251$ & $0.1847$\\
$\bar{B}_c^{*}$ & $0.3040$ & $0.3099$ & $0.3133$ & $0.3896$ & $0.3099$ & $0.2489$\\
$\eta_c$        & $0.0537$ & $0.0537$ & $0.0537$ & $0.0663$ & $0.0537$ & $0.0441$\\
$J/\psi$        & $0.0556$ & $0.0555$ & $0.0555$ & $0.0686$ & $0.0555$ & $0.0456$\\
\hline\hline
\end{tabular}
\label{tab:unc_masses}
\end{table}

Therefore, within the framework of four-body phase space of these top-quark decay channels in this work, the dominant mass-induced theoretical uncertainty is more likely to be $m_c$,  because the heavy charm mass shrinks the available phase space and thereby suppresses the production rate of mesons. On the other hand,  improve the precision of $m_c$ together with the related non-perturbative long-distance matrix element, such as the wave functions at the origin, will be most effective for sharpening the phenomenological predictions.

\begin{figure*}[htbp]
\centering
\subfigure{\includegraphics[width=0.28\textwidth]{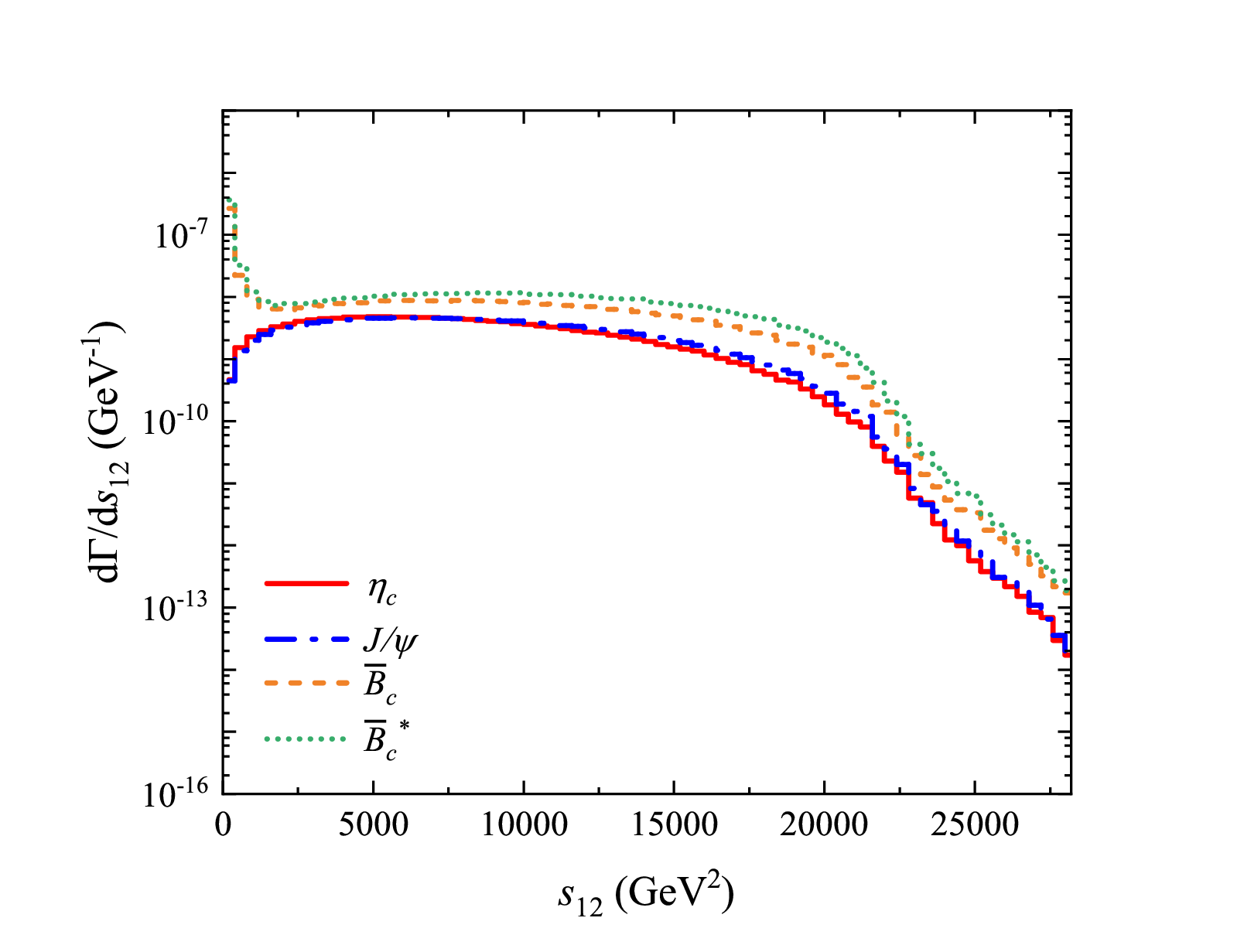}}\hspace{-0.05\textwidth}
\subfigure{\includegraphics[width=0.28\textwidth]{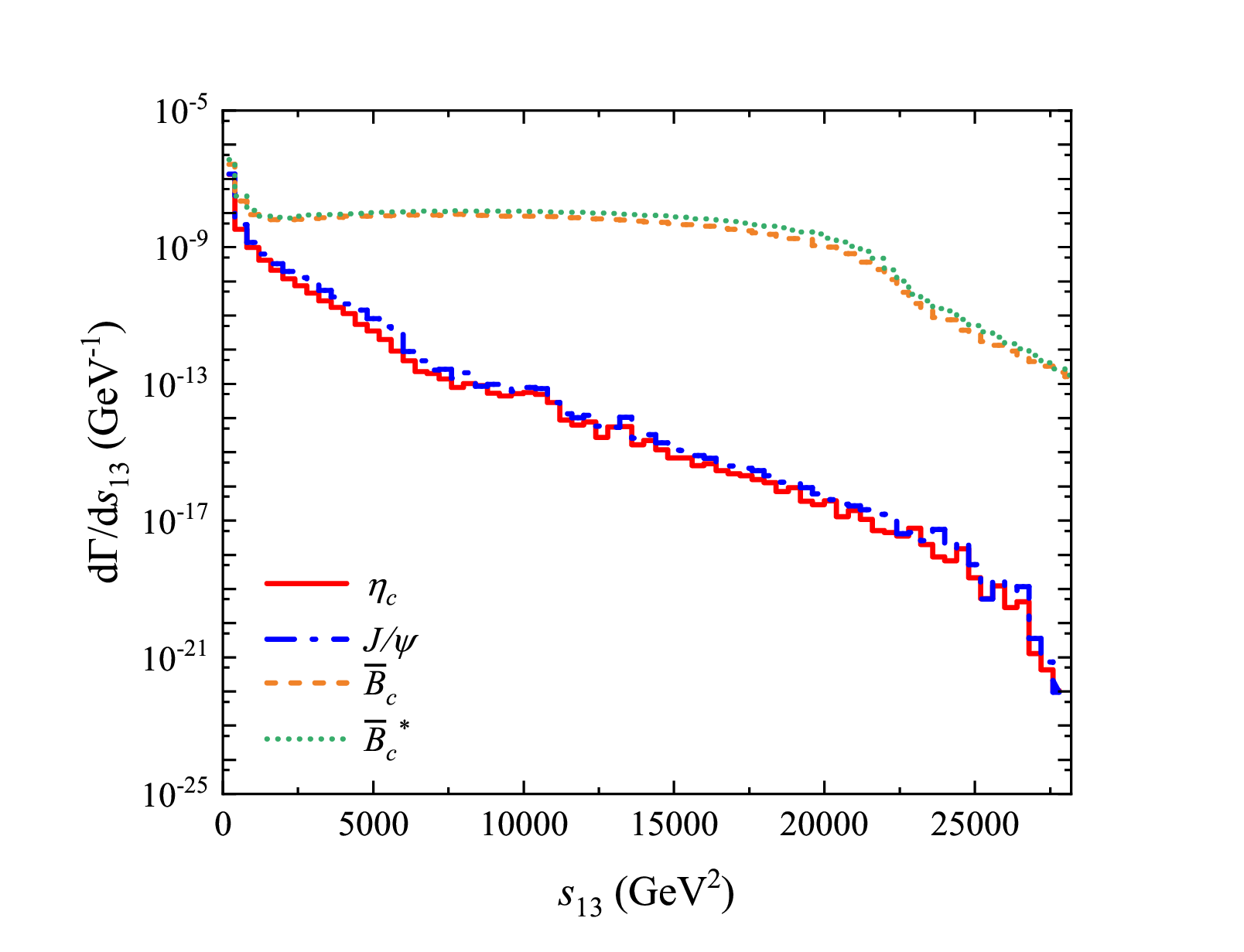}}\hspace{-0.05\textwidth}
\subfigure{\includegraphics[width=0.28\textwidth]{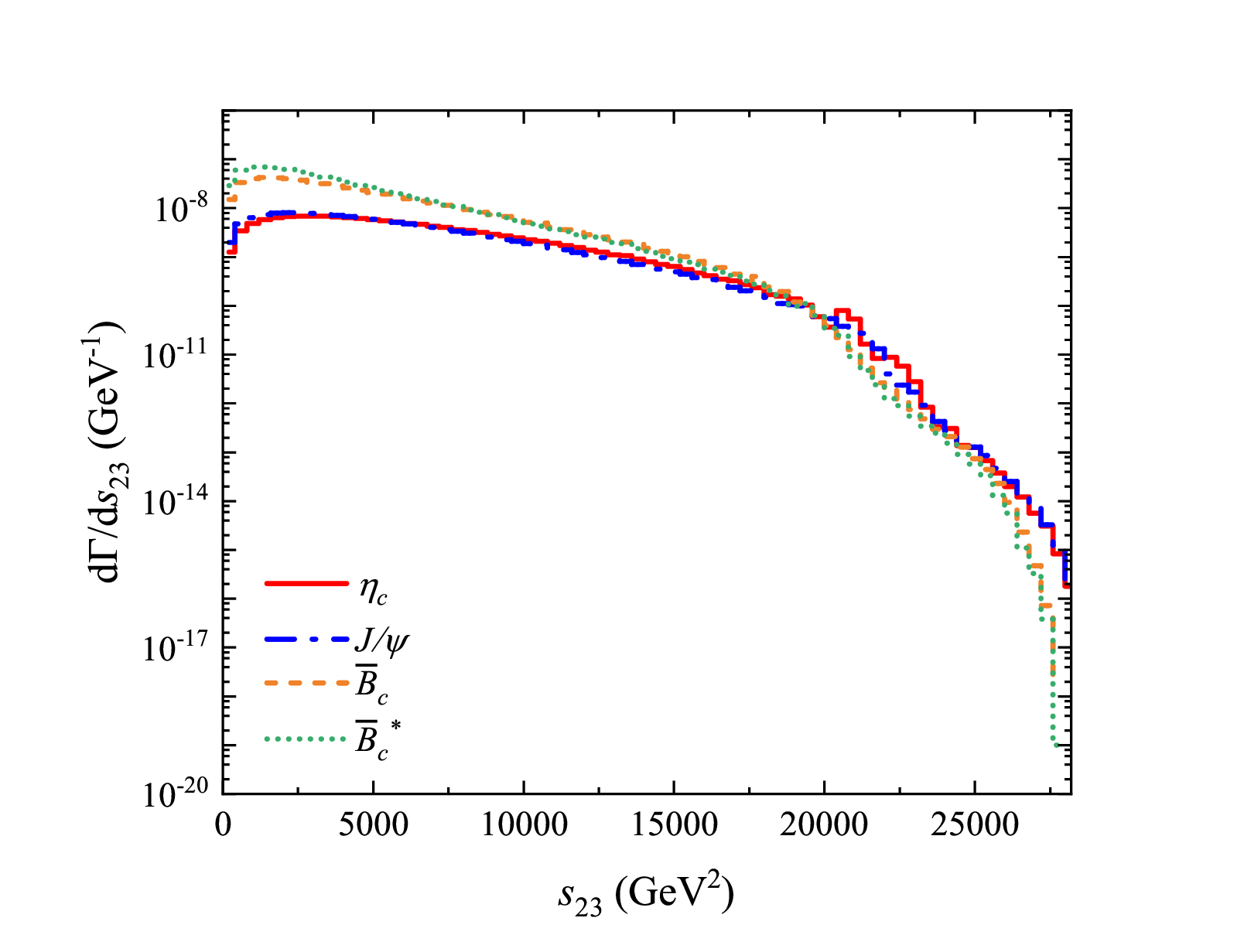}}\hspace{-0.05\textwidth}
\subfigure{\includegraphics[width=0.28\textwidth]{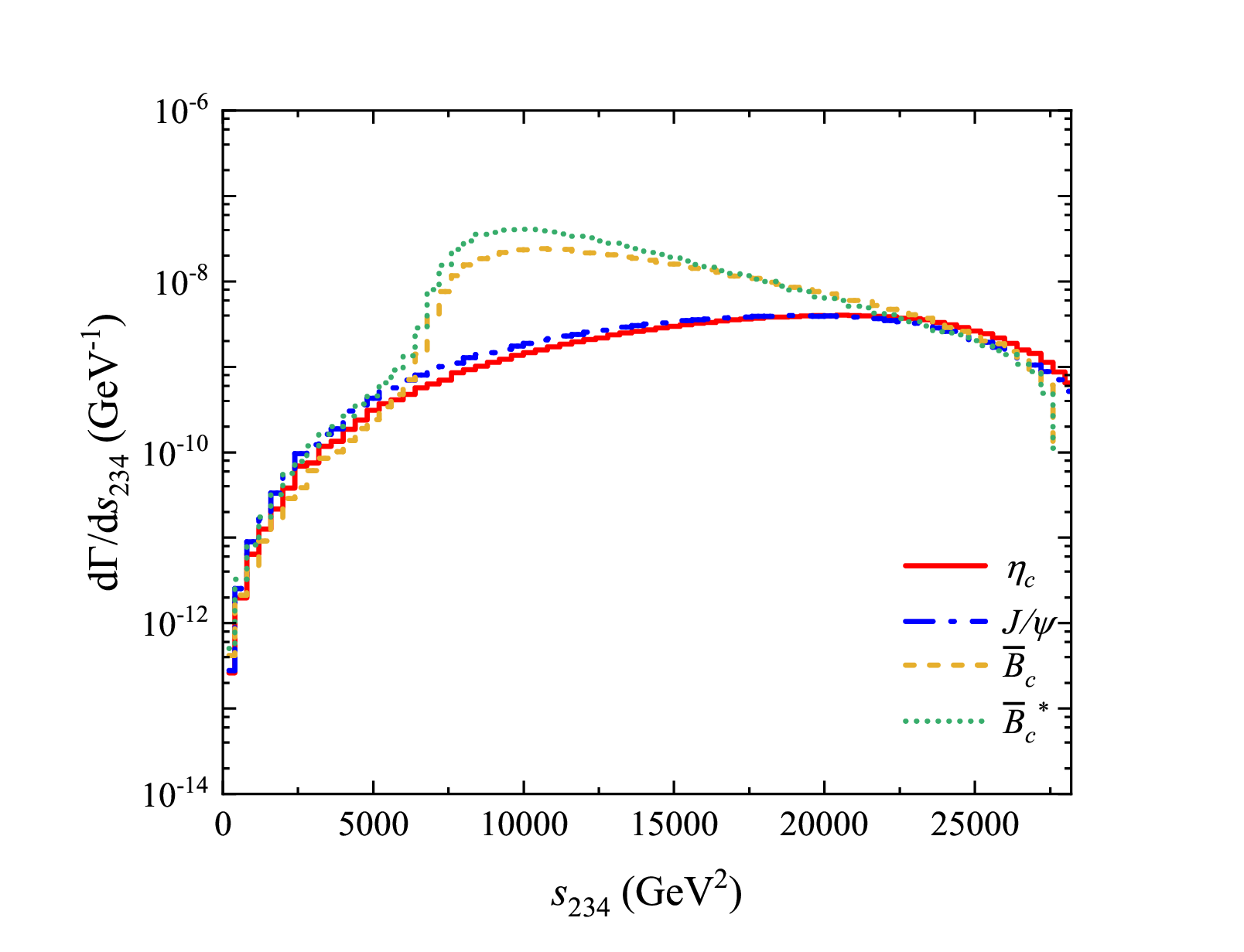}}
\caption{Invariant-mass differential distributions for the four-body top-quark decay channels.
From left to right, the panels show $d\Gamma/ds_{12}$, $d\Gamma/ds_{13}$, $d\Gamma/ds_{23}$, and $d\Gamma/ds_{234}$, with $s_{ij}=(p_i+p_j)^2$ and $s_{234}=(p_2+p_3+p_4)^2$, for the $(b\bar{c})$ and $(c\bar{c})$ quarkonium production.}
\label{fig:compare_s}
\end{figure*}

\begin{figure*}[htbp]
\centering
\subfigure{\includegraphics[width=0.28\textwidth]{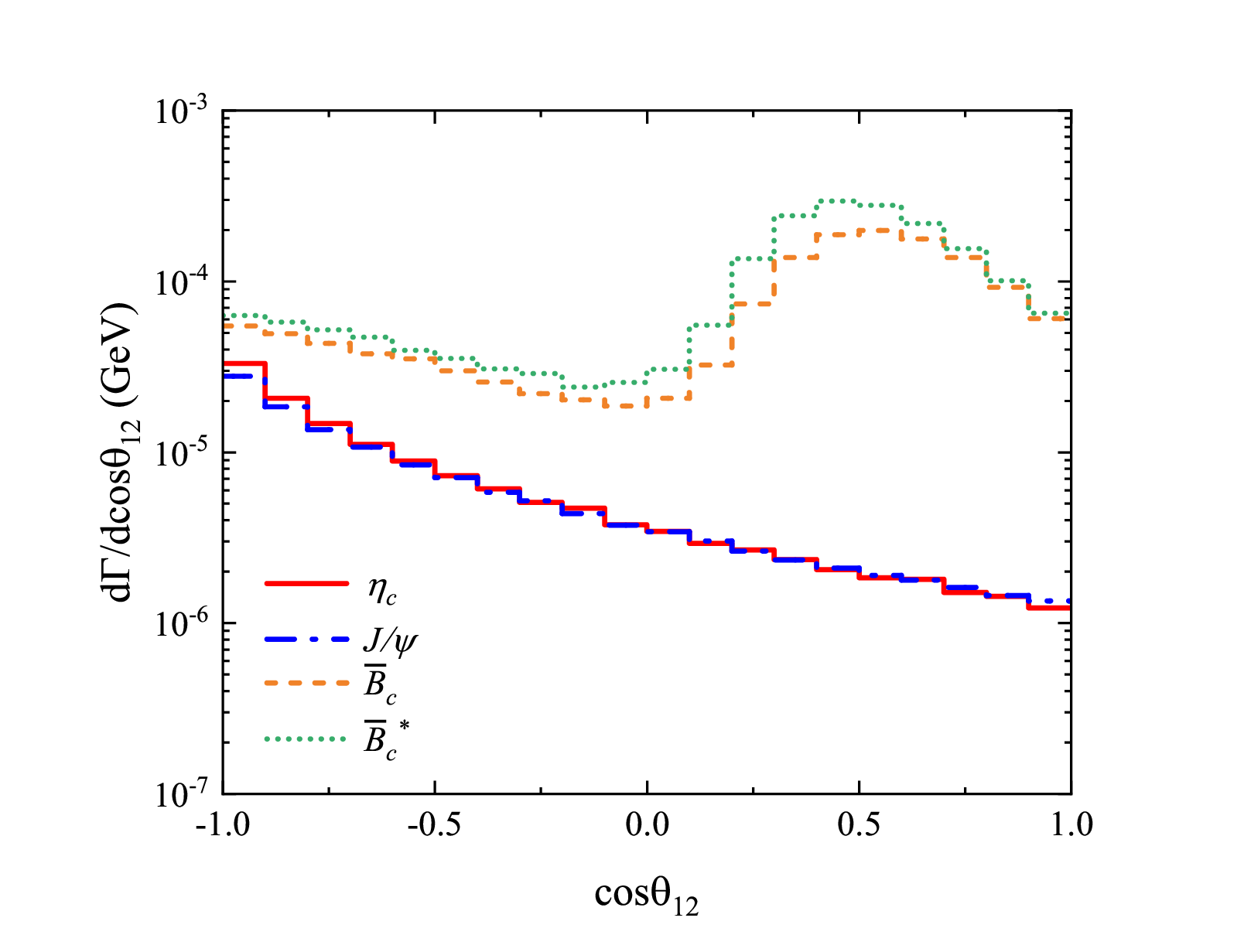}}\hspace{-0.05\textwidth}
\subfigure{\includegraphics[width=0.28\textwidth]{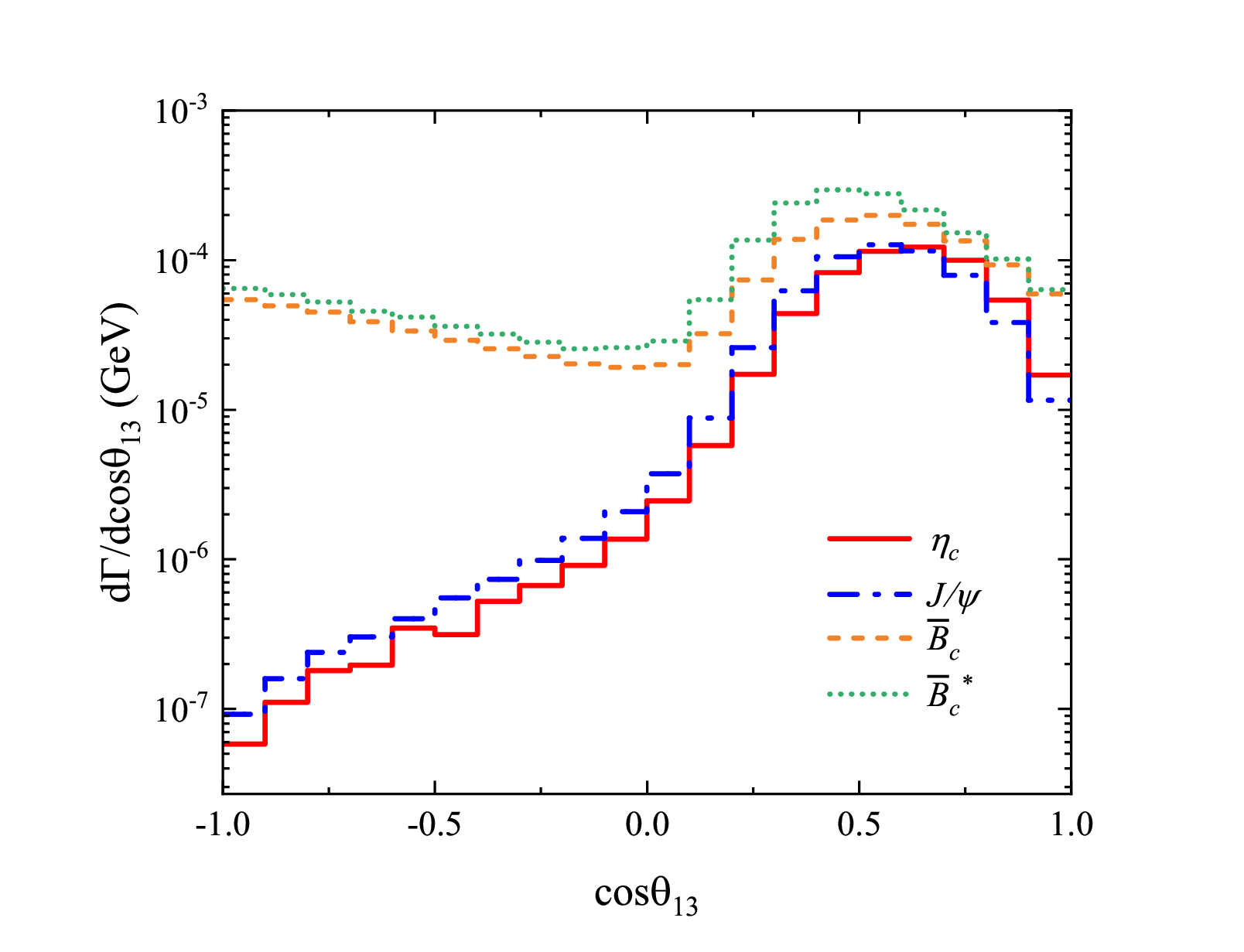}}\hspace{-0.05\textwidth}
\subfigure{\includegraphics[width=0.28\textwidth]{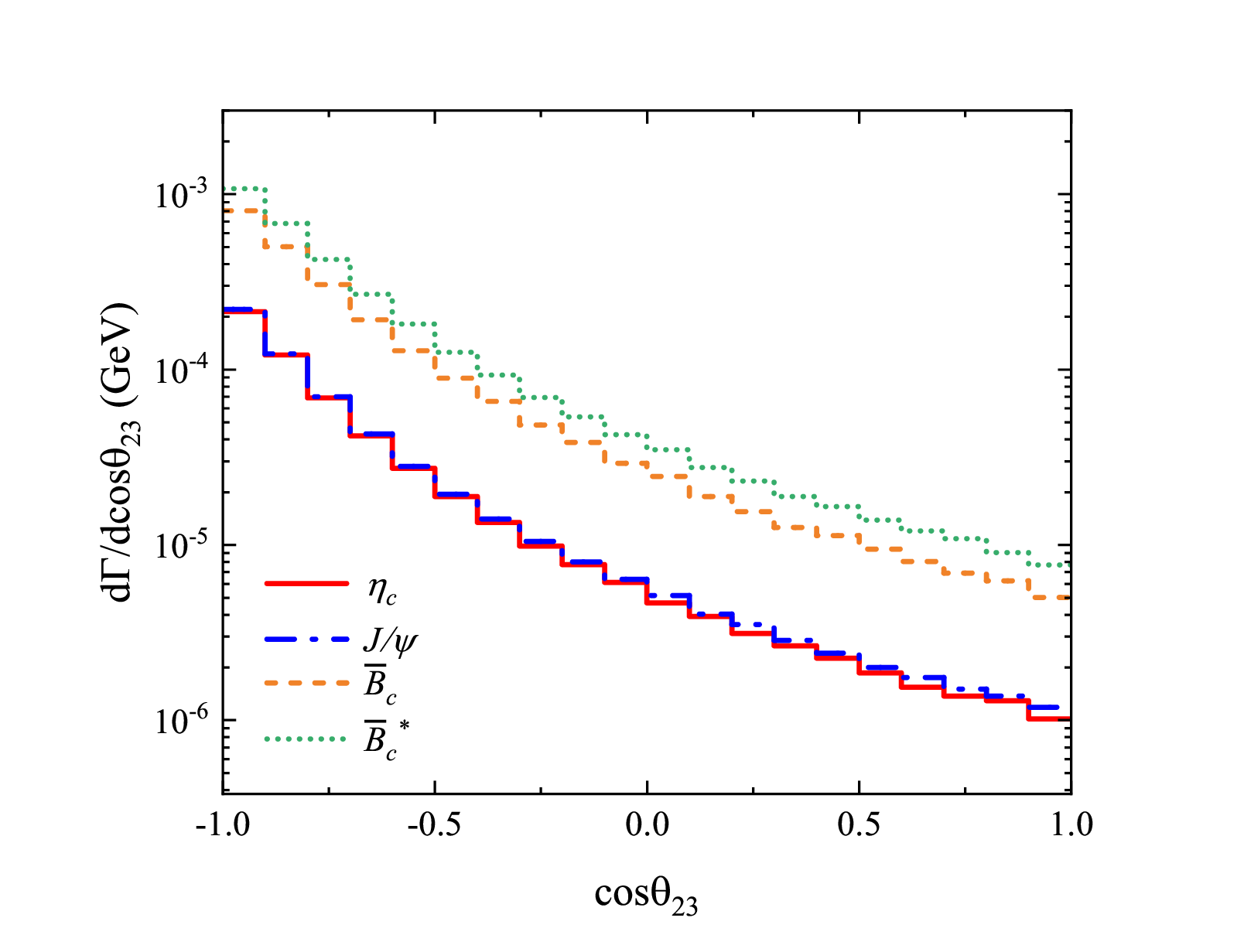}}\hspace{-0.05\textwidth}
\subfigure{\includegraphics[width=0.28\textwidth]{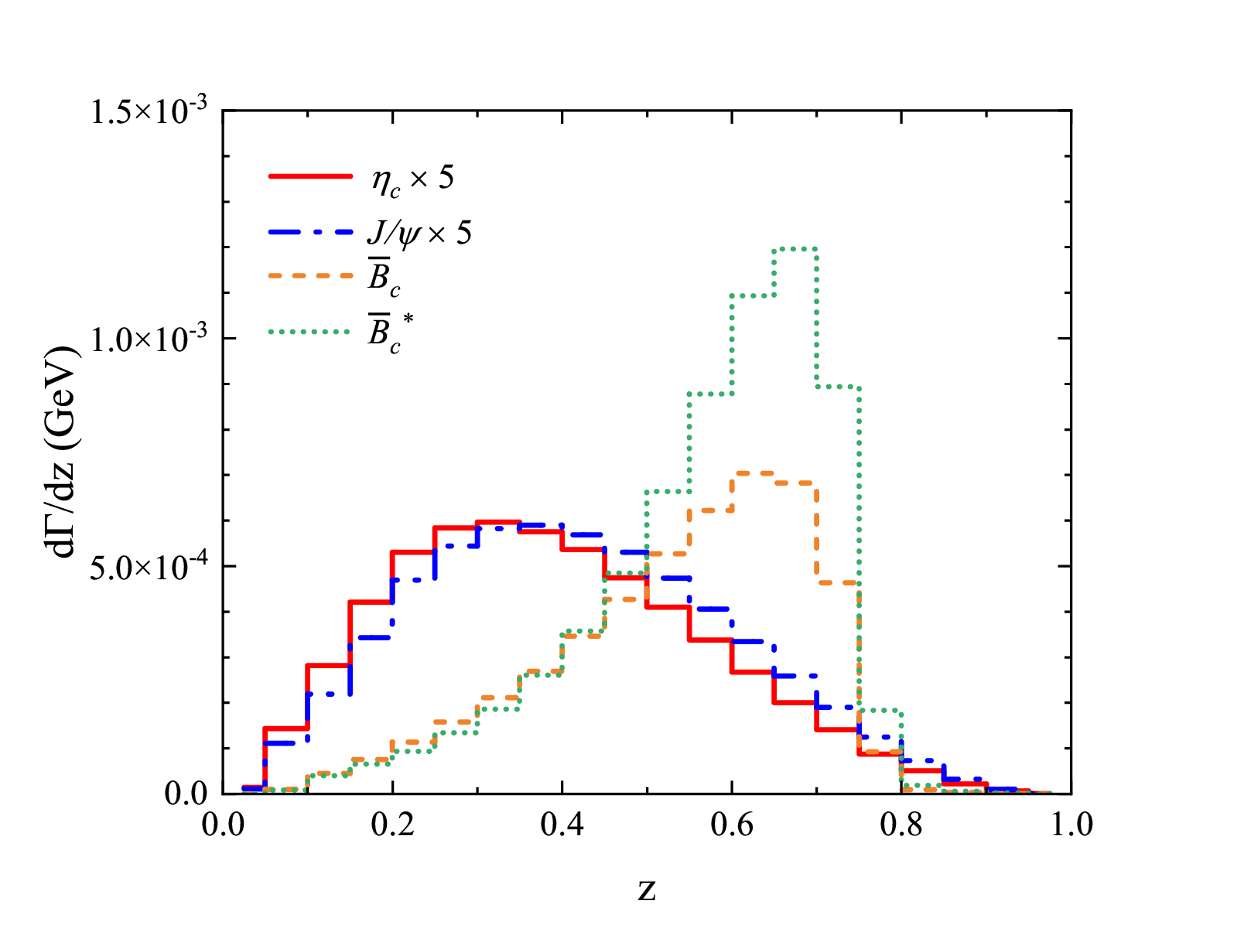}}
\caption{Angular and $z$ differential distributions for the four-body top-quark decay channels. 
From left to right, the panels show $d\Gamma/d\cos\theta_{12}$, $d\Gamma/d\cos\theta_{13}$, $d\Gamma/d\cos\theta_{23}$, and $d\Gamma/dz$, where $\theta_{ij}$ denotes the angle between $\vec p_i$ and $\vec p_j$ in the top-quark rest frame, and $z$ is the normalized longitudinal momentum fraction of the produced doubly heavy quarkonium. }
\label{fig:compare_costh}
\end{figure*}

Finally, in order to provide more kinematic information for future experimental searches, the differential distributions of invariant masses, angles, and $z$ are presented in Fig.~\ref{fig:compare_s} and Fig.~\ref{fig:compare_costh}, respectively. Here, the squared invariant masses are defined as $s_{ij}=(p_i + p_j)^2$ and $s_{ijk}=(p_i + p_j +p_k)^2$. $\theta_{ij}$ is the angle between momenta $\vec{p}_i$ and $\vec{p}_j$ in the top-quark rest frame. $z$ is the longitudinal momentum fraction of the $c\bar{c}$ ($b\bar{c}$)
relative to the $W^+$ boson ($b$ quark), i.e. $z=E_{c\bar{c}}/E_{c\bar{c}}^{\max}$ ($E_{b\bar{c}}/E_{b\bar{c}}^{\max}$). As shown in Fig.~\ref{fig:compare_s}, the invariant-mass distributions of $s_{12}$, $s_{13}$, and $s_{23}$ all generally show a downward trend. The $[^3S_1]$ and $[^1S_0]$ states of $(b\bar{c})$ and $(c\bar{c})$ show similar trends and are difficult to distinguish. The invariant-mass distributions of $s_{234}$ have a mountain-like shape, and especially the $b\bar{c}$ quarkonium have a much steeper profile. The $\cos\theta_{23}$ distribution in Fig.~\ref{fig:compare_costh} indicates that when $\theta_{23}=\pi$ (i.e., when the two particles $b$ ($c$) and $c$ are back-to-back for the production of $(b\bar{c})$ quarkonium ($(c\bar{c})$ quarkonium), the largest contribution is obtained. 

However, for the production of $(b\bar{c})$ and $(c\bar{c})$ quarkonium, there are significant differences in the angular distributions of $\cos\theta_{12}$ and $\cos\theta_{13}$. For $B_c$-meson production, $t\to (b\bar{c})+c+c+\bar{s}$, the two charm quarks in the final state are identical. They can either come from gluon splitting radiated off the $b$ jet (more likely to be collinear with the $(b\bar{c})$ quarkonium) or from the $W$-boson decay (more likely to be opposite to the $(b\bar{c})$ quarkonium). Therefore, the $B_c$ meson has a large probability to be either aligned with or anti-aligned with $p_2$ (equivalently $p_3$), which implies that the angular distributions of $\cos\theta_{12}$ and $\cos\theta_{13}$ should be the same and non-monotonic. For $c\bar{c}$ meson production, $t\to (c\bar{c})+ b + c +\bar{s}$, the $(c\bar{c})$ quarkonium is mainly produced from $W^+$-boson decays, so it is more likely to be back-to-back with the $b$ jet, leading to a monotonic decrease in the $\cos\theta_{12}$ distribution. Meanwhile, due to the identical charm quarks in the final state, the $\cos\theta_{13}$ distribution remains non-monotonic. 
The $z$ distributions of the $b\bar{c}$  and $c\bar{c}$ mesons also show significant differences. This is mainly because the $b$ quark is much heavier than the $c$ quark and therefore ($b\bar{c}$)-quarkonium carries a larger fraction of the parent momentum.

\noindent\textbf{\textit{Summary}---}
In this work, we study a four-body ($1\to4$) top-quark decay mechanism to produce a doubly heavy quarkonium associated with two heavy quarks and one light antiquark,
$t \to (b\bar{c}) + c + c + \bar{s}$ and $t \to (c\bar{c}) + b + c + \bar{s}$, within the NRQCD factorization framework.
The decay widths are factorized into perturbatively calculable short-distance coefficients and non-perturbative long-distance matrix elements.
At leading order, the hard amplitudes are obtained from the tree-level weak decay chain $t\to bW^+$ and $W^+\to c\bar{s}$ accompanied by an extra hard gluon splitting into a $c\bar{c}$ pair; the quarkonium formation is implemented by projecting the corresponding heavy-quark pair onto definite spin and color states.
We focus on the dominant color-singlet $S$-wave channels $(b\bar{c})[^1S_0]_1$, $(b\bar{c})[^3S_1]_1$, $(c\bar{c})[^1S_0]_1$, and $(c\bar{c})[^3S_1]_1$, and perform the four-body phase-space integration using a recursive parameterization in terms of invariant masses and angles.

Numerically, for this newly identified channel we obtain the partial widths
$\Gamma(t\to \bar{B}_c + c + c + \bar{s})=0.2251~\mathrm{MeV}$,
$\Gamma(t\to \bar{B}_c^{*} + c + c + \bar{s})=0.3099~\mathrm{MeV}$,
$\Gamma(t\to \eta_c + b + c + \bar{s})=0.0537~\mathrm{MeV}$, and
$\Gamma(t\to J/\psi + b + c + \bar{s})=0.0555~\mathrm{MeV}$.
An important phenomenological observation is that, among the considered top-quark decay mechanisms, the four-body channel studied here provides a dominant contribution to $J/\psi$ and $\eta_c$ production.
Moreover, this high-energy $1\to4$ process provides an additional and useful testing ground for examining the validity and applicability of the NWA.
To facilitate experimental investigations at top-rich facilities, we also present a detailed discussion of theoretical uncertainties and provide differential distributions (such as angular and invariant-mass spectra), which encode characteristic kinematic features induced by the heavy-quark masses and the multi-body phase space.

\noindent\textbf{Acknowledgments:} This work was supported in
part by the Natural Science Foundation of China Grant No. 12505106 and No.12547101, and by the Natural Science Foundation of Guangxi Grant No. 2024GXNSFBA010368 and No. 2025GXNSFAA069775. H.-H. Ma also acknowledges support from Funda\c{c}\~ao de Amparo \`a Pesquisa do Estado de S\~ao Paulo (FAPESP), Brazil, Process No. 2025/01276-7. \\


\begin{thebibliography}{99}

\bibitem{E598:1974sol}
J.~J.~Aubert \textit{et al.} [E598],
Phys. Rev. Lett. \textbf{33}, 1404-1406 (1974).

\bibitem{SLAC-SP-017:1974ind}
J.~E.~Augustin \textit{et al.} [SLAC-SP-017],
Phys. Rev. Lett. \textbf{33}, 1406-1408 (1974).

\bibitem{CDF:1998axz}
F.~Abe \textit{et al.} [CDF],
Phys. Rev. D \textbf{58}, 112004 (1998).

\bibitem{Bodwin:1994jh}
G.~T.~Bodwin, E.~Braaten and G.~P.~Lepage,
Phys. Rev. D \textbf{51}, 1125-1171 (1995)
[erratum: Phys. Rev. D \textbf{55}, 5853 (1997)].

\bibitem{Petrelli:1997ge}
A.~Petrelli, M.~Cacciari, M.~Greco, F.~Maltoni and M.~L.~Mangano,
Nucl. Phys. B \textbf{514}, 245 (1998).

\bibitem{Braaten:1993jn}
E.~Braaten, K.~m.~Cheung and T.~C.~Yuan,
Phys. Rev. D \textbf{48}, R5049 (1993).

\bibitem{Zheng:2019gnb}
X.~C.~Zheng, C.~H.~Chang, T.~F.~Feng and X.~G.~Wu,
Phys. Rev. D \textbf{100}, 034004 (2019).

\bibitem{Chang:1992jb}
C.~H.~Chang and Y.~Q.~Chen,
Phys. Rev. D \textbf{48}, 4086-4091 (1993).

\bibitem{Chang:1994aw}
C.~H.~Chang, Y.~Q.~Chen, G.~P.~Han and H.~T.~Jiang,
Phys. Lett. B \textbf{364}, 78-86 (1995).

\bibitem{Gershtein:1994jw}
S.~S.~Gershtein, V.~V.~Kiselev, A.~K.~Likhoded and A.~V.~Tkabladze,
Phys. Usp. \textbf{38}, 1-37 (1995).

\bibitem{Berezhnoy:1994ba}
A.~V.~Berezhnoy, A.~K.~Likhoded and M.~V.~Shevlyagin,
Phys. Atom. Nucl. \textbf{58}, 672-689 (1995).

\bibitem{Chang:2003cr}
C.~H.~Chang and X.~G.~Wu,
Eur. Phys. J. C \textbf{38}, 267-276 (2004).

\bibitem{Chang:2004bh}
C.~H.~Chang, J.~X.~Wang and X.~G.~Wu,
Phys. Rev. D \textbf{70}, 114019 (2004).

\bibitem{Chang:2005bf}
C.~H.~Chang, C.~F.~Qiao, J.~X.~Wang and X.~G.~Wu,
Phys. Rev. D \textbf{71}, 074012 (2005).

\bibitem{Zheng:2015ixa}
X.~C.~Zheng, C.~H.~Chang and Z.~Pan,
Phys. Rev. D \textbf{93}, 034019 (2016).

\bibitem{Zhang:2021ypo}
Z.~Y.~Zhang, X.~C.~Zheng and X.~G.~Wu,
Eur. Phys. J. C \textbf{82}, 246 (2022).

\bibitem{Yang:2022zpc}
Z.~Yang, X.~C.~Zheng and X.~G.~Wu,
Comput. Phys. Commun. \textbf{281}, 108503 (2022).

\bibitem{Zhan:2022etq}
X.~J.~Zhan, X.~G.~Wu and X.~C.~Zheng,
Phys. Rev. D \textbf{106}, 094036 (2022).

\bibitem{Chen:2014frw}
G.~Chen, X.~G.~Wu, Z.~Sun, Y.~Ma and H.~B.~Fu,
JHEP \textbf{12}, 018 (2014).

\bibitem{Sun:2015hhv}
Z.~Sun, X.~G.~Wu and H.~F.~Zhang,
Phys. Rev. D \textbf{92}, 074021 (2015).

\bibitem{Bi:2016vbt}
H.~Y.~Bi, R.~Y.~Zhang, H.~Y.~Han, Y.~Jiang and X.~G.~Wu,
Phys. Rev. D \textbf{95}, 034019 (2017).

\bibitem{Zhan:2022nck}
X.~J.~Zhan, X.~G.~Wu and X.~C.~Zheng,
JHEP \textbf{09}, 050 (2022).

\bibitem{Cai:2026hll}
N.~Cai, X.~J.~Zhan and T.~F.~Feng,
[arXiv:2602.05348 [hep-ph]].

\bibitem{Chen:2018obq}
G.~Chen, C.~H.~Chang and X.~G.~Wu,
Phys. Rev. D \textbf{97}, 114022 (2018).

\bibitem{Qiao:1996rd}
C.~F.~Qiao, C.~S.~Li and K.~T.~Chao,
Phys. Rev. D \textbf{54}, 5606-5610 (1996).

\bibitem{Chang:2007si}
C.~H.~Chang, J.~X.~Wang and X.~G.~Wu,
Phys. Rev. D \textbf{77}, 014022 (2008).

\bibitem{Wu:2008cn}
X.~G.~Wu,
Phys. Lett. B \textbf{671}, 318-322 (2009).

\bibitem{Sun:2010rw}
P.~Sun, L.~P.~Sun and C.~F.~Qiao,
Phys. Rev. D \textbf{81}, 114035 (2010).

\bibitem{Chang:1992bb}
C.~H.~Chang and Y.~Q.~Chen,
Phys. Rev. D \textbf{46}, 3845 (1992)
[erratum: Phys. Rev. D \textbf{50}, 6013 (1994)].

\bibitem{Deng:2010aq}
L.~C.~Deng, X.~G.~Wu, Z.~Yang, Z.~Y.~Fang and Q.~L.~Liao,
Eur. Phys. J. C \textbf{70}, 113-124 (2010).

\bibitem{Yang:2010yg}
Z.~Yang, X.~G.~Wu, L.~C.~Deng, J.~W.~Zhang and G.~Chen,
Eur. Phys. J. C \textbf{71}, 1563 (2011).

\bibitem{Wang:2023ssg}
G.~Y.~Wang, X.~C.~Zheng, X.~G.~Wu and G.~Z.~Xu,
Phys. Rev. D \textbf{109}, 074004 (2024).

\bibitem{Zheng:2022mds}
X.~C.~Zheng, X.~G.~Wu, X.~J.~Zhan, H.~Zhou and H.~T.~Li,
Phys. Rev. D \textbf{106}, 094008 (2022).

\bibitem{Qiao:2011zc}
C.~F.~Qiao, L.~P.~Sun and R.~L.~Zhu,
JHEP \textbf{08}, 131 (2011).

\bibitem{Liao:2011kd}
Q.~L.~Liao, X.~G.~Wu, J.~Jiang, Z.~Yang and Z.~Y.~Fang,
Phys. Rev. D \textbf{85}, 014032 (2012).

\bibitem{Liao:2012rh}
Q.~L.~Liao, X.~G.~Wu, J.~Jiang, Z.~Yang, Z.~Y.~Fang and J.~W.~Zhang,
Phys. Rev. D \textbf{86}, 014031 (2012).

\bibitem{Qiao:2011yk}
C.~F.~Qiao, L.~P.~Sun, D.~S.~Yang and R.~L.~Zhu,
Eur. Phys. J. C \textbf{71}, 1766 (2011)
doi:10.1140/epjc/s10052-011-1766-3
[arXiv:1103.1106 [hep-ph]].

\bibitem{Zheng:2019egj}
X.~C.~Zheng, C.~H.~Chang, X.~G.~Wu, J.~Zeng and X.~D.~Huang,
Phys. Rev. D \textbf{101}, 034029 (2020).

\bibitem{Jiang:2015pah}
J.~Jiang and C.~F.~Qiao,
Phys. Rev. D \textbf{93}, 054031 (2016).

\bibitem{Zheng:2023atb}
X.~C.~Zheng, X.~G.~Wu, X.~J.~Zhan, G.~Y.~Wang and H.~T.~Li,
Phys. Rev. D \textbf{107}, 074005 (2023).

\bibitem{Niu:2018tvo}
J.~J.~Niu, L.~Guo, H.~H.~Ma and S.~M.~Wang,
Eur. Phys. J. C \textbf{78}, 657 (2018).

\bibitem{Chang:2003cq}
C.~H.~Chang, C.~Driouichi, P.~Eerola and X.~G.~Wu,
Comput. Phys. Commun. \textbf{159}, 192-224 (2004).

\bibitem{Chang:2005hq}
C.~H.~Chang, J.~X.~Wang and X.~G.~Wu,
Comput. Phys. Commun. \textbf{174}, 241-251 (2006).

\bibitem{ParticleDataGroup:2024cfk}
S.~Navas \textit{et al.} [Particle Data Group],
Phys. Rev. D \textbf{110}, 030001 (2024).

\bibitem{Breit:1936zzb}
G.~Breit and E.~Wigner,
Phys. Rev. \textbf{49}, 519-531 (1936)

\bibitem{Jackson:1975vf}
J.~D.~Jackson and D.~L.~Scharre,
Nucl. Instrum. Meth. \textbf{128}, 13 (1975)


\bibitem{Uhlemann:2009}
C.~F.~Uhlemann and N.~Kauer,
Nucl. Phys. B \textbf{814}, 195-211 (2009).

\bibitem{Heinrich:2018}
G.~Heinrich, A.~Maier, R.~Nisius, J.~Schlenk, M.~Schulze, L.~Scyboz and J.~Winter,
JHEP \textbf{07}, 129 (2018).



\bibitem{Eichten:1994gt}
E.~J.~Eichten and C.~Quigg,
Phys. Rev. D \textbf{49}, 5845-5856 (1994).

\bibitem{Eichten:1995ch}
E.~J.~Eichten and C.~Quigg,
Phys. Rev. D \textbf{52}, 1726-1728 (1995).


\bibitem{Kidonakis:2004hr}
N.~Kidonakis and R.~Vogt,
Int. J. Mod. Phys. A \textbf{20}, 3171 (2005).

\bibitem{Kuhn:2013zoa}
J.~H.~K{\"u}hn, A.~Scharf and P.~Uwer,
Phys. Rev. D \textbf{91}, 014020 (2015).

\bibitem{CEPCStudyGroup:2023quu}
W.~Abdallah \textit{et al.} [CEPC Study Group],
Radiat. Detect. Technol. Methods \textbf{8}, no.1, 1-1105 (2024)
[erratum: Radiat. Detect. Technol. Methods \textbf{9}, no.1, 184-192 (2025)]

\bibitem{Ai:2024nmn}
X.~Ai, W.~Altmannshofer, P.~Athron, X.~Bai, L.~Calibbi, L.~Cao, Y.~Che, C.~Chen, J.~Y.~Chen and L.~Chen, \textit{et al.}
Chin. Phys. \textbf{49}, no.10, 103003 (2025)


\bibitem{LHeC:2020van}
P.~Agostini \textit{et al.} [LHeC and FCC-he Study Group],
J. Phys. G \textbf{48}, no.11, 110501 (2021)


\end{thebibliography}
\end{document}